\documentclass{article}

\usepackage{amsfonts}
\usepackage{amsmath}
\usepackage{amsthm}
\usepackage{arxiv}
\usepackage{booktabs}
\usepackage{doi}
\usepackage[T1]{fontenc}
\usepackage{graphicx}
\usepackage{hyperref}
\usepackage[utf8]{inputenc}
\usepackage{mathrsfs}
\usepackage{microtype}
\usepackage{nicefrac}
\usepackage{url}
\usepackage{xcolor}

\theoremstyle{plain}

\newtheorem{lemma}{Lemma}
\newtheorem{theorem}{Theorem}

\theoremstyle{definition}

\theoremstyle{remark}

\title{Shock-Wave Refinement of the Friedmann--Robertson--Walker Metric}

\author{ \href{https://orcid.org/0000-0001-9255-6281}{\includegraphics[scale=0.06]{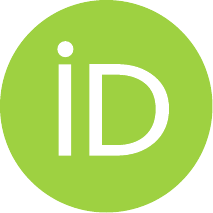}\hspace{1mm}Christopher E.~Alexander}\\
Department of Mathematics\\
University College London\\
London, WC1H 0AY\\
United Kingdom\\
\texttt{christopher.alexander@ucl.ac.uk}\\
\And
\href{https://orcid.org/0000-0002-6907-1101}{\includegraphics[scale=0.06]{orcid.pdf}\hspace{1mm}Blake Temple} \\
Department of Mathematics\\
University of California\\
Davis, CA 95616\\
United States\\
\texttt{temple@math.ucdavis.edu}\\
\And
{Joel A.~Smoller}\thanks{Joel Smoller passed away in 2017 and contributed to the original version of this article, which was supported in part by NSF Applied Mathematics Grant Number DMS-010-3998.}\\
Department of Mathematics\\
University of Michigan\\
Ann Arbor, MI 48109\\
United States\\
}

\renewcommand{\headeright}{}
\renewcommand{\undertitle}{}
\renewcommand{\shorttitle}{Shock-Wave Refinement of the Friedmann--Robertson--Walker Metric}

\hypersetup{
pdftitle={Shock-Wave Refinement of the Friedmann--Robertson--Walker Metric},
pdfsubject={gr-qc, math-ph},
pdfauthor={Christopher E.~Alexander, Blake Temple, Joel A.~Smoller},
pdfkeywords={General Relativity, Einstein-Euler, Self-Similar, Shock Wave, Cosmology, Dark Energy},
}

\begin{document}

\maketitle

\begin{abstract}
	The mathematics of general relativistic shock waves is introduced and considered in a cosmological context. In particular, an expanding Friedmann--Roberson--Walker metric is matched to a Tolman--Oppenheimer--Volkoff metric across a spherical shock surface. This is the general relativistic analogue of a shock-wave explosion within a static singular isothermal fluid sphere and may be regarded as a model for the Big Bang. These shock waves are constructed both within and beyond the Hubble radius, which corresponds to a universe outside and inside its Schwarzschild radius respectively. Certain self-similar perturbations of the FRW metric lead to an accelerated expansion, even without a cosmological constant, and thus it is conjectured that such a mechanism may account for the anomalous acceleration observed today without recourse to dark energy.
\end{abstract}


\vfill

\tableofcontents

\vfill

\pagebreak

\section{Introduction}
\label{S1}

In the Standard Model of Cosmology, the expanding universe of galaxies is described by a Friedmann--Robertson--Walker (FRW) metric, which in spherical coordinates has a line element given by,
\begin{equation}
	ds^2 = -dt^2 + R^2(t)\left(\frac{dr^2}{1-kr^2} + r^2[d\theta^2 + \sin^2(\theta)d\phi^2]\right).\label{FRW}
\end{equation}
In this model, which accounts for physics on the largest length scale, the Universe is approximated by a space of uniform density and pressure at each fixed time, and the expansion rate is determined by the cosmological scale factor $R(t)$ that evolves according to the Einstein field equations. Astronomical observations indicate that the galaxies are uniform on a scale of about one billion light-years, and the expansion is critical, that is, $k=0$ in (\ref{FRW}). According to (\ref{FRW}) with $k=0$, on the largest scale the Universe is infinitely flat Euclidean space $\mathbb{R}^3$ at each fixed time. Matching the Hubble constant to its observed values and invoking the Einstein field equations, the FRW model implies that the entire infinite universe $\mathbb{R}^3$ emerged all at once from a singularity ($R=0$) some 13.7 billion years ago, and this event is referred to as the \emph{Big Bang}.

This article, updated in 2023, summarises the work of Smoller and Temple in \cite{ST1994,ST1995,ST2003,ST2004}, and then describes subsequent advances by Smoller, Temple and Alexander in \cite{ST2009,ST2012,A2022}. The Smoller--Temple solutions describe a two-parameter family of exact solutions to the perfect fluid Einstein field equations that refine the FRW metric by a spherical shock-wave cutoff. The Einstein field equations with a perfect fluid source are also referred to as the \emph{Einstein-Euler} equations and we make the distinction that an \emph{exact} solution is one that is defined by a system of ODE, whereas an \emph{explicit} solution is one that can be given in terms of elementary mathematical functions. In the original Smoller--Temple solutions, the flat ($k=0$) FRW spacetime approximates the expanding wave created behind the general relativistic version of an explosion into a static singular isothermal sphere. There are two cases: matching \emph{outside the black hole}, which means the shock surface is inside the Hubble radius of the FRW spacetime, and \emph{inside the black hole}, that is, beyond the Hubble radius. In the examples given in the original article, Smoller and Temple recognised that the flat FRW spacetime did not fully resolve the expanding wave behind the shock, but argued that it was qualitatively close. After Smoller and Temple introduced a family of expanding waves in \cite{ST2012}, Alexander in \cite{A2022} accomplished the goal of fully resolving the expanding wave behind these general relativistic shock waves for the case of shock waves inside the Hubble radius, that is, outside the black hole. This is accomplished by incorporating self-similar perturbations of the FRW spacetime, characterised in \cite{ST2012} as the natural self-similar extension of the FRW spacetime, that is, spacetimes that are asymptotically FRW at the origin. Alexander's solution for pure radiation resolves a long-standing open problem first proposed by Taub in \cite{CT1971}, and later taken up from a different point of view by Smoller and Temple. Most interestingly, the expanding waves behind these shocks naturally introduce accelerations relative to the flat FRW spacetime. Since the flat FRW spacetime is the starting model of Cosmology, it is intriguing to consider these models as an explanation for cosmic acceleration without recourse to a cosmological constant. We argue here that the analysis needs to be extended to a parallel family of solutions beyond the Hubble radius, as accomplished in \cite{ST2003,ST2004} for flat FRW shock matching, to make a plausible case for such an explanation. We first review these Smoller--Temple solutions and then return in Section \ref{S6} to discuss the refinements made by Smoller, Temple and Alexander.

In order to construct a mathematically simple family of shock-wave refinements of the FRW metric that satisfy the Einstein-Euler equations exactly, we assume critical expansion ($k=0$) and restrict to the case that the sound speed in the fluid on the FRW side of the shock wave is constant, that is, we assume an FRW equation of state $p=\sigma\rho$ with $0 < \sigma \leq c^2$, where $\sigma$, the square of the sound speed $\sqrt{\frac{\partial p}{\partial\rho}}$, is constant. For $\sigma = c^2/3$, this equation of state describes a state of matter known as \emph{pure radiation}, as well as the equation of state of the relativistic limit of free particles, which is correct during the Radiation Dominated Epoch of the Early Universe \cite{W1972}. Also, as $\sigma$ ranges from $0$ to $c^2$, we obtain qualitatively correct approximations to general equations of state. Now by using units such that the speed of light $c$ and gravitational constant $\mathcal{G}$ are set to unity, the family of solutions is then determined by two parameters: $0 < \sigma \leq 1$ and $r_* \geq 0$. The second parameter $r_*$ is the FRW radial coordinate $r$ of the shock in the limit $t\rightarrow0$, that is, the instant of the Big Bang.\footnotemark[1]\footnotetext[1]{Since when $k=0$ the FRW metric is invariant under the rescaling $r\rightarrow\alpha r$ and $R\rightarrow R/\alpha$, we fix the radial coordinate $r$ by fixing the scale factor $\alpha$ with the condition that $R(t_0)=1$ for some time $t_0$, say present time.} The FRW radial coordinate $r$ is singular with respect to radial arc-length $\bar{r} = rR$ at the Big Bang ($R=0$), so setting $r_*>0$ does not place the shock wave away from the origin at time $t=0$. The distance from the FRW centre to the shock wave tends to zero in the limit $t\rightarrow0$, even when $r_*>0$. In the limit $r_*\rightarrow\infty$, we recover from the family of solutions the usual (infinite) FRW metric with equation of state $p = \sigma\rho$, that is, we recover the standard FRW metric in the limit that the shock wave is infinitely far out. In this sense, our family of solutions of the Einstein-Euler equations represents a two-parameter refinement of the standard Friedmann--Robertson--Walker metric.

The explicitly defined solutions in the case $r_*=0$ were first constructed in \cite{ST1995} and are qualitatively different from the exact solutions in the case $r_*>0$, which were constructed later in \cite{ST2004}. The difference is that when $r_*=0$, the shock wave lies closer than one Hubble length from the centre of the FRW spacetime throughout its motion, but when $r_*>0$, the shock wave emerges at the Big Bang at a distance beyond one Hubble length \cite{ST2000}. The Hubble length, also referred to as the Hubble radius in certain contexts, depends on time, and tends to zero as $t\rightarrow0$. We show in \cite{ST2004} that one Hubble length, equal to $c/H$ where $H = \dot{R}/R$, is a critical length scale in a flat FRW metric because the total mass inside one Hubble length has a Schwarzschild radius equal to exactly one Hubble length.\footnotemark[2]\footnotetext[2]{Since $c/H$ is a good estimate for the age of the Universe, it follows that the Hubble length $c/H$ is approximately the distance of light travel starting at the Big Bang up until present time. In this sense, the Hubble length is a rough estimate for the distance to the further most objects visible in the Universe.} That is, one Hubble length marks precisely the distance at which the Schwarzschild radius $\bar{r}_s \equiv 2M$ (of the mass $M$ inside a radial shock wave at distance $\bar{r}$ from the FRW centre) crosses from inside ($\bar{r}_s<\bar{r}$) to outside ($\bar{r}_s>\bar{r}$) the shock wave. If the shock wave is at a distance closer than one Hubble length from the FRW centre, then $\bar{r} > 2M$ and we say that the solution lies \emph{outside the black hole}, but if the shock wave is at a distance greater than one Hubble length, then $\bar{r} < 2M$ at the shock, and we say the solution lies \emph{inside the black hole}. Since $M$ increases proportional to $\bar{r}^3$, it follows that $\bar{r} > 2M$ for $\bar{r}$ sufficiently small, and $\bar{r} < 2M$ for $\bar{r}$ sufficiently large, so there must be a critical radius at which $\bar{r} = 2M$. In Section 2 (taken from \cite{ST2003,ST2004}), we show that when $k=0$, this critical radius is exactly the Hubble radius. When the parameter $r_*=0$, the family of solutions for $0 < \sigma \leq 1$ starts at the Big Bang, and evolves thereafter outside the black hole, satisfying $\bar{r} > 2M$ everywhere from $t=0$ onward. But when $r_*>0$, the shock wave is further out than one Hubble length at the instant of the Big Bang, and the solution begins with $\bar{r} < 2M$ at the shock wave. From this time onward, the spacetime expands until eventually the Hubble radius catches up to the shock wave at $\bar{r} = 2M$ and then passes the shock wave, making $\bar{r} > 2M$ thereafter. Thus when $r_*>0$, the whole spacetime begins inside the black hole (with $\bar{r} < 2M$ for sufficiently large $\bar{r}$) but eventually evolves to a solution outside the black hole. The time when $\bar{r} = 2M$ actually marks the event horizon of a \emph{white hole}, the time reversal of a black hole, in the ambient spacetime beyond the shock wave. We show that when $r_*>0$, the time when the Hubble length catches up to the shock wave comes before the time when the shock wave comes into view at the FRW centre. Furthermore, when $\bar{r} = 2M$, assuming $t$ is so large that we can neglect the pressure from this time onward, the whole solution emerges from the white hole as a finite ball of mass expanding into empty space, satisfying $\bar{r} > 2M$ everywhere thereafter. In fact, when $r_*>0$, the zero pressure Oppenheimer--Snyder solution outside the black hole gives the large time asymptotics of the solution (see \cite{OS1939,ST1998A,ST2004} and the comments after Theorems \ref{T6}-\ref{T8} below).

The explicitly defined solutions in the case $r_*=0$ give a general relativistic version of an explosion into a static singular isothermal fluid sphere, qualitatively similar to the corresponding classical explosion outside the black hole \cite{ST1995}. The main difference physically between the cases $r_*>0$ and $r_*=0$ is that when $r_*>0$, that is, the case when the shock wave emerges from the Big Bang at a distance beyond one Hubble length, a large region of uniform expansion is created behind the shock wave at the instant of the Big Bang. Thus, when $r_*>0$, lightlike (also known as \emph{null}) information about the shock wave propagates inward from the wave, rather than outward from the centre, as is the case when $r_*=0$ and the shock lies inside one Hubble length.\footnotemark[3]\footnotetext[3]{One can imagine that when $r_*>0$, the shock wave could get out through a great deal of matter early on when everything is dense and compressed and still not violate the speed of light bound. Thus when $r_*>0$, the shock wave would \emph{thermalise}, or more accurately, \emph{make uniform} a large region at the centre early on in the explosion. The authors speculate that such a mechanism might offer a substitute for inflation in this regard.} It follows that when $r_*>0$, an observer positioned in the FRW spacetime inside the shock wave will see exactly what the Standard Model of Cosmology predicts up until the time when the shock wave comes into view in the far field. In this sense, the case $r_*>0$ gives a black hole cosmology that refines the standard FRW model of Cosmology to the case of finite mass. One of the surprising differences between the case $r_*=0$ and the case $r_*>0$ is that, when $r_*>0$, the important pure radiation equation of state $p = \frac{1}{3}\rho$ comes out of the analysis as special at the Big Bang. When $r_*>0$, the shock wave emerges at the instant of the Big Bang at a finite non-zero speed (the speed of light) only for the special value $\sigma = \frac{1}{3}$. In this case, the equation of state on both sides of the shock wave tends to the correct relation $p = \frac{1}{3}\rho$ as $t\rightarrow0$ and the shock wave decelerates to a subluminal speed for all positive times thereafter (see \cite{ST2003,ST2004} and Theorem \ref{T8} below).

In all cases $0<\sigma\leq1$, $r_*\geq0$, the spacetime metric that lies beyond the shock wave is taken to be a metric of Tolman--Oppenheimer--Volkoff (TOV) form \cite{OV1939}, that is,
\begin{equation}
	ds^2 = -B(\bar{r})d\bar{t}^2 + \frac{1}{A(\bar{r})}d\bar{r}^2 + \bar{r}^2[d\theta^2 + \sin^2(\theta)d\phi^2].\label{TOV}
\end{equation}
The metric (\ref{TOV}) is in \emph{standard Schwarzschild coordinates}, that is, diagonal with a radial coordinate defined by spheres of symmetry. Furthermore, the metric components depend only on the radial coordinate $\bar{r}$. Barred coordinates are used to distinguish TOV coordinates from unbarred FRW coordinates for the purpose of matching the metric at the shock surface (see Section \ref{S3}). The mass function $M(\bar{r})$ enters as a metric component through the relation,
\begin{equation}
	A = 1 - \frac{2M(\bar{r})}{\bar{r}}.
\end{equation}
The TOV metric (\ref{TOV}) has a very different character depending on whether $A>0$ or $A<0$, that is, depending on whether the solution lies outside or inside the black hole. In the case $A>0$, $\bar{r}$ is a spacelike coordinate and the TOV metric describes a static general relativistic fluid sphere.\footnotemark[4]\footnotetext[4]{The metric (\ref{TOV}) is, for example, the starting point for the stability limits of Buchdahl and Chandrasekhar for stars \cite{W1972,ST1997A,ST1998B}.} When $A<0$, $\bar{r}$ is the timelike coordinate and (\ref{TOV}) is a dynamic metric that evolves in time. The explicitly defined shock-wave solutions are obtained by taking $\bar{r} = R(t)r$ to match the spheres of symmetry, and then matching the metrics (\ref{FRW}) and (\ref{TOV}) at an interface $\bar{r} = \bar{r}(t)$ across which the metrics are Lipschitz continuous. This can be done in general. In order for the interface to be a physically meaningful shock surface, we use the result in Theorem \ref{T4} below (see \cite{ST1994}), so that a single additional conservation constraint is sufficient to rule out delta function sources at the shock and guarantee that the matched metric solves the Einstein-Euler equations in the weak sense.\footnotemark[5]\footnotetext[5]{The Einstein field equations $G = \kappa T$ are second-order in the metric and so delta function sources will in general be present at a Lipschitz continuous matching of metrics.} The Lipschitz continuous matching of the metrics, together with the conservation constraint, leads to a system of ordinary differential equations (ODE) that determine the shock position, together with the TOV density and pressure at the shock. Since the TOV metric depends only on $\bar{r}$, the equations thus determine the TOV spacetime beyond the shock wave. To obtain a physically meaningful outgoing shock wave, we impose the constraint $\bar{p} \leq \bar{\rho}$ to ensure that the equation of state on the TOV side of the shock is qualitatively reasonable, and we require that the shock be compressive as the entropy condition. For an outgoing shock wave, this corresponds to the conditions $\rho>\bar{\rho}$ and $p>\bar{p}$, that is, the pressure and density need to be larger on the side of the shock that receives the mass flux (the FRW side when the shock wave is propagating away from the FRW centre). This condition breaks the time reversal symmetry of the equations and is sufficient to rule out rarefaction shocks in classical gas dynamics \cite{S1983,ST2004}. The ODE, together with the equation of state bound and the conservation and entropy constraints, determine a unique solution of the ODE for every $0 < \sigma \leq 1$ and $\bar{r}_*\geq0$, and this provides the two-parameter family of solutions discussed here \cite{ST1995,ST2004}.

The Lipschitz matching of the metrics implies that the total mass $M$ is continuous across the interface, and so when $r_*>0$, the total mass of the entire solution (inside and outside the shock wave) is finite at each time $t>0$. Both the FRW and TOV spacetimes emerge at the Big Bang. The total mass $M$ on the FRW side of the shock has the meaning of total mass inside radius $\bar{r}$ at fixed time, but on the TOV side of the shock, $M$ does not evolve according to equations that give it the interpretation as a total mass because the metric is inside the black hole. Nevertheless, after the spacetime emerges from the black hole, the total mass takes on its usual meaning outside the black hole and time asymptotically the Big Bang ends with an expansion of finite total mass in the usual sense. Thus, when $r_*>0$, our shock-wave refinement of the FRW metric leads to a Big Bang of finite total mass.

The Smoller--Temple family of shock-wave solutions are rough models in the sense that the equation of state on the FRW side has constant $\sigma$ and the equation of state on the TOV side is determined by the equations and therefore cannot be imposed. For more accurate equations of state, a more accurate description of the expanding wave created behind the shock is needed to meet the conservation constraint and thereby mediate the transition across the shock wave. At the time of publish of the original article, the authors thought such expanding waves to be pretty much impossible to model as exact solutions, however, in the most recent work of Alexander \cite{A2022}, this has been resolved for shock waves inside the Hubble radius by considering self-similar perturbations of the flat FRW spacetime. Not only do these modifications permit enough parameter freedom to impose the same equation of state on each side of the shock, but the perturbations of the flat FRW spacetime induce an accelerated expansion without the presence of a cosmological constant. A rigorous proof for the existence of a general relativistic shock wave inside the Hubble radius with a pure radiation equation of state each side of the shock is also demonstrated in \cite{A2022}. The fact that we can find global solutions that meet our physical bounds within and beyond the Hubble radius, that are qualitatively the same for all values of $\sigma\in (0,1]$ and all initial shock positions, strongly suggests that such a shock wave would be the dominant wave in a large class of problems.

In Section \ref{S2} we derive the FRW solution for constant $\sigma$ and discuss the Hubble radius as a critical length scale. In Section \ref{S3} we state the general theorems in \cite{ST1994} for matching gravitational metrics across shock waves. In Section \ref{S4} we discuss the construction of the family of solutions in the case $r_*=0$. In Section \ref{S5} we discuss the case $r_*>0$, and in Section \ref{S6} we discuss the advances made since 2005, in particular, replacing the flat FRW solution with a family of self-similar perturbations and discussing how such waves exhibit an accelerated expansion. See \cite{ST1995,ST2004,ST2012,A2022} for details.

\section{The FRW Metric}
\label{S2}

According to Einstein's Theory of General Relativity, all properties of the gravitational field are determined by a Lorentzian spacetime metric tensor $g$, whose line element in a given coordinate system $\boldsymbol{x} = (x^0,...,x^3)$ is given by
\begin{equation}
	ds^2 = g_{\mu\nu}dx^\mu dx^\nu.
\end{equation}
We use the Einstein summation convention whereby repeated up-down indices are assumed to be summed from 0 to 3. The components $g_{\mu\nu}$ of the gravitational metric $g$ satisfy the Einstein field equations,
\begin{align}
	G^{\mu\nu} &= \kappa T^{\mu\nu}, & T^{\mu\nu} &= (\rho c^2 + p)u^\mu u^\nu + pg^{\mu\nu},\label{EE}
\end{align}
where we assume the stress-energy-momentum tensor $T$ is that of a perfect fluid, making these equations the \emph{Einstein-Euler} equations. Here $G$ is the Einstein curvature tensor,
\begin{equation}
	\kappa = \frac{8\pi\mathcal{G}}{c^4}\label{kappa}
\end{equation}
is the coupling constant, $\mathcal{G}$ is Newton's gravitational constant, $c$ is the speed of light, $\rho c^2$ is the energy density, $p$ is the pressure and $\boldsymbol{u} = (u^0,...,u^3)$ is the fluid four-velocity \cite{W1972}. We will also use the convention $c=1$ and $\mathcal{G}=1$ when convenient.

Putting the metric ansatz (\ref{FRW}) into the Einstein-Euler equations (\ref{EE}) gives the equations for the components of FRW metric \cite{W1972},
\begin{align}
	H^2 = \left(\frac{\dot{R}}{R}\right)^2 &= \frac{\kappa}{3}\rho-\frac{k}{R^2},\label{FRW1}\\
	\dot{\rho} &= -3(p+\rho)H.\label{FRW2}
\end{align}
The unknowns $R$, $\rho$ and $p$ are assumed to be functions of the FRW coordinate time $t$ alone, with the dot denoting differentiation with respect to $t$. To verify that the Hubble radius $\bar{r}_{crit} = 1/H$ is the limit for FRW-TOV shock matching outside a black hole, write the FRW metric (\ref{FRW}) in standard Schwarzschild coordinates $\bar{\boldsymbol{x}} = (\bar{t},\bar{r})$ where the metric takes the form
\begin{equation}
	ds^2 = -B(\bar{t},\bar{r})d\bar{t}^2 + \frac{1}{A(\bar{t},\bar{r})}d\bar{r}^2 + \bar{r}^2d\Omega^2,\label{SSchwarz}
\end{equation}
and the mass function $M(\bar{t},\bar{r})$ is defined through the relation
\begin{equation}
	A = 1 - \frac{2M}{\bar{r}}.\label{M1}
\end{equation}
It is well known that a general spherically symmetric metric can be put in the form (\ref{SSchwarz}) by coordinate transformation \cite{W1972,GT2004}. Substituting $\bar{r}=Rr$ into (\ref{FRW}) and diagonalising the resulting metric (see \cite{ST2004} for details), we obtain
\begin{equation}
	ds^2 = \frac{1}{1-kr^2-H^2\bar{r}^2}\left(-\frac{1-kr^2}{\psi^2}d\bar{t}^2 + d\bar{r}^2\right) + \bar{r}^2d\Omega^2,\label{FRWSchwarz1}
\end{equation}
where $\psi$ is an integrating factor that solves the equation
\begin{equation}
	\frac{\partial}{\partial \bar{r}}\left(\psi\frac{1-kr^2-H^2\bar{r}^2}{1-kr^2}\right) - \frac{\partial}{\partial t}\left(\psi \frac{H\bar{r}}{1-kr^2}\right) = 0,\label{FRWSchwarz2}
\end{equation}
and the time coordinate $\bar{t} = \bar{t}(t,\bar{r})$ is defined by the exact differential
\begin{equation}
	d\bar{t} = \left(\psi\frac{1-kr^2-H^2\bar{r}^2}{1-kr^2}\right)dt + \left(\psi\frac{H\bar{r}}{1-kr^2}\right)d\bar{r}.\label{FRWSchwarz3}
\end{equation}
Now using (\ref{M1}) in (\ref{FRW1}), it follows that
\begin{equation}
	M(t,\bar{r}) = \frac{\kappa}{2}\int_0^{\bar{r}}\rho(t)s^2ds = \frac{1}{3}\frac{\kappa}{2}\rho\bar{r}^3.\label{M2}
\end{equation}
Since in the FRW metric $\bar{r}=Rr$ measures arc-length along radial geodesics at fixed time, we see from (\ref{M2}) that $M(t,\bar{r})$ has the physical interpretation as the total mass inside radius $\bar{r}$ at time $t$ in the FRW metric. Restricting to the case of critical expansion ($k=0$), we see from (\ref{FRW1}), (\ref{FRWSchwarz3}) and (\ref{M2}) that $\bar{r} = 1/H$ is equivalent to $\bar{r} = 2M$, and so at fixed time $t$, the following equivalences are valid:
\begin{equation}
	\bar{r} = \frac{1}{H} \iff \frac{2M}{\bar{r}} = 1 \iff A = 0.
\end{equation}
We conclude that $\bar{r} = 1/H$ is the critical length scale for the FRW metric at fixed time $t$ in the sense that $A$ changes sign at $\bar{r} = 1/H$, and so the universe lies inside a black hole beyond $\bar{r} = 1/H$, as claimed above. It is shown in \cite{ST1998B} that the standard TOV metric outside the black hole cannot be continued into $A=0$ except in the very special case $\rho=0$, as it takes an infinite pressure to hold up a static configuration at the event horizon of a black hole. Thus to do shock matching beyond one Hubble length requires a metric of a different character, and for this purpose, in \cite{ST2003,ST2004} we introduce the TOV metric \emph{inside the black hole}, that is, a metric of TOV form with $A<0$ and whose fluid is co-moving with the timelike radial coordinate $\bar{r}$.

The Hubble radius $\bar{r}_{crit} = c/H$ is also the critical distance at which the outward expansion of the FRW metric exactly cancels the inward advance of a radial light ray impinging on an observer positioned at the origin of a flat ($k=0$) FRW metric. Indeed, by (\ref{FRW}), a light ray travelling radially inward toward the centre of an FRW coordinate system satisfies,
\begin{equation}
	c^2dt^2 = R^2dr^2,\label{LightRay}
\end{equation}
so that
\begin{equation}
	\frac{d\bar{r}}{dt} = \dot{R}r+R\dot{r}=H\bar{r}-c = H\left(\bar{r}-\frac{c}{H}\right)>0,
\end{equation}
if and only if $\bar{r} > c/H$. Thus the arc-length distance from the origin to an inward moving light ray at fixed time $t$ in a flat FRW metric will actually increase as long as the light ray lies beyond the Hubble radius. An inward moving light ray will, however, eventually cross the Hubble radius and reach the origin in finite proper time due to the increase in the Hubble length with time.

We now calculate the infinite redshift limit in terms of the Hubble length. It is well known that light emitted at $(t_e,r_e)$ at wavelength $\lambda_e$ in an FRW spacetime will be observed at $(t_0,r_0)$ at wavelength $\lambda_0$ if
\begin{equation}
	\frac{R_0}{R_e} = \frac{\lambda_0}{\lambda_e}.
\end{equation}
Moreover, the redshift factor $z$ is defined by
\begin{equation}
	z = \frac{\lambda_0}{\lambda_e}-1.
\end{equation}
Thus, infinite redshift occurs in the limit $R_e\rightarrow0$, where $R=0$, $t=0$ is the Big Bang. Consider now a light ray emitted at the instant of the Big Bang and observed at the FRW origin at present time $t=t_0$. Let $r_\infty$ denote the FRW coordinate at time $t\rightarrow 0$ of the furthermost objects that can be observed at the FRW origin before time $t=t_0$. Then $r_\infty$ marks the position of objects at time $t=0$ whose radiation would be observed as infinitely redshifted (assuming no scattering). Note then that a shock wave emanating from $\bar{r}=0$ at the instant of the Big Bang will be observed at the FRW origin before present time $t=t_0$ only if its position $r$ at the instant of the Big Bang satisfies $r<r_\infty$. To estimate $r_\infty$, note first that from (\ref{LightRay}) it follows that an incoming radial light ray in an FRW metric follows a null trajectory $r=r(t)$ if
\begin{equation}
	r-r_e = -\int_{t_e}^{t}\frac{d\tau}{R(\tau)},
\end{equation}
and thus
\begin{equation}
	r_\infty=\int_{0}^{t_0}\frac{d\tau}{R(\tau)}.
\end{equation}
Using this, the following theorem is proved in \cite{ST2004}.
\begin{theorem}
	If the pressure $p$ satisfies the bounds
	\begin{equation}
		0\leq p\leq\frac{1}{3}\rho,
	\end{equation}
	then for any equation of state, the age of the Universe $t_0$ and the infinite red shift limit $r_\infty$ are bounded in terms of the Hubble length by:
	\begin{align}
		\frac{1}{2H_0} \leq &t_0 \leq \frac{2}{3H_0},\label{Bounds}\\
		\frac{1}{H_0} \leq &r_\infty \leq \frac{2}{H_0}.\label{rInfinityBound}
	\end{align}
	Where we have assumed that $R=0$ when $t=0$ and $R=1$ when $t=t_0$, $H=H_0$.
\end{theorem}
The next theorem gives closed form solutions of the FRW equations (\ref{FRW1}), (\ref{FRW2}) for constant $\sigma$. As a special case, we recover the bounds in (\ref{Bounds}) and (\ref{rInfinityBound}) from the cases $\sigma=0$ and $\sigma = \frac{1}{3}$.
\begin{theorem}
	Assume $k=0$ and the equation of state $p = \sigma\rho$, where $0 \leq \sigma \leq 1$ is constant. Then, for an expanding FRW universe ($\dot{R}>0$), the solution of system (\ref{FRW1}), (\ref{FRW2}) satisfying $R=0$ at $t=0$ and $R=1$ at $t=t_0$ is given by,
	\begin{align}
		\rho &= \frac{4}{3\kappa(1+\sigma)^2}\frac{1}{t^2},\label{FRW3}\\
		R &= \left(\frac{t}{t_0}\right)^{\frac{2}{3(1+\sigma)}},\label{FRW4}\\
		\frac{H}{H_0} &= \frac{t_0}{t}.\label{FRW5}
	\end{align}
	Moreover, the age of the Universe $t_0$ and the infinite redshift limit $r_\infty$ are given explicitly in terms of the Hubble length by:
	\begin{align}
		t_0 &= \frac{2}{3(1+\sigma)}\frac{1}{H_0},\\
		r_\infty &= \frac{2}{1+3\sigma}\frac{1}{H_0}.\label{rInfinityExplicit}
	\end{align}
\end{theorem}
From (\ref{rInfinityExplicit}) we conclude that a shock wave will be observed at the FRW origin before present time $t=t_0$ only if its position $r$ at the instant of the Big Bang satisfies $r < r_\infty$. Note that $r_\infty$ ranges from one half to two Hubble lengths as $\sigma$ ranges from $1$ to $0$, taking the intermediate value of one Hubble length at $\sigma = \frac{1}{3}$. Note also that by using (\ref{FRW3})-(\ref{FRW4}) in (\ref{M2}), it follows that
\begin{equation}
	M = \frac{\kappa}{2}\int_0^{\bar{r}}\rho(t)s^2ds = \frac{2\bar{r}^3}{9(1+\sigma)^2t_0^{\frac{2}{1+\sigma}}}\frac{1}{t^{\frac{2\sigma}{1+\sigma}}},
\end{equation}
so $\dot{M}<0$ if $\sigma>0$. It follows that if $p = \sigma\rho$, and $\sigma$ is a positive constant, then the total mass inside a radius of constant $r$ decreases in time.

\section{The General Theory of Shock Matching}
\label{S3}

The matching of the FRW and TOV metrics in the next two sections is based on the following theorems that are derived in \cite{ST1994}.\footnotemark[6]\footnotetext[6]{Theorems \ref{T3} and \ref{T4} apply to non-null shock surfaces.}
\begin{theorem}
	\label{T3}
	Let $\Sigma$ denote a smooth three-dimensional shock surface with spacelike normal vector $\boldsymbol{n}$ relative to the spacetime metric $g$, let $K$ denote the second fundamental form on $\Sigma$ and let $G$ denote the Einstein curvature tensor. Assume that the components $g_{\mu\nu}$ of the gravitational metric $g$ are continuous up to the boundary on either side separately and Lipschitz continuous across $\Sigma$ in some fixed coordinate system. Then the following statements are equivalent:
	\begin{enumerate}
		\item $[K]=0$ at each point of $\Sigma$.
		\item The curvature tensors $R^\mu_{\nu\sigma\tau}$ and $G_{\mu\nu}$, viewed as second-order operators on the metric components $g_{\mu\nu}$, produce no delta function sources on $\Sigma$.
		\item For each point $P\in\Sigma$ there exists a $C^{1,1}$ coordinate transformation defined in a neighbourhood of $P$, such that, in the new coordinates (which can be taken to be the Gaussian normal coordinates for the surface), the metric components are $C^{1,1}$ functions of these coordinates.
		\item For each $P\in\Sigma$, there exists a coordinate frame that is locally Lorentzian at $P$, and can be reached within the class of $C^{1,1}$ coordinate transformations.
	\end{enumerate} 
	Moreover, if any one of these equivalencies hold, then the Rankine--Hugoniot jump conditions,
	\begin{equation}
		[G^{\mu\nu}]n_\mu = 0\label{RH1}
	\end{equation}
	hold at each point on $\Sigma$.
\end{theorem}
The Rankine--Hugoniot jump conditions express the weak form of conservation of energy and momentum across $\Sigma$ when $G = \kappa T$. Here $[f]$ denotes the jump in the quantity $f$ across $\Sigma$, which is determined by the metric separately on each side of $\Sigma$ since $g_{\mu\nu}$ is only Lipschitz continuous across $\Sigma$. The notation $C^{1,1}$ denotes a function whose first derivatives are Lipschitz continuous. In the case of spherical symmetry, a stronger result holds. In this case, the jump conditions (\ref{RH1}) are implied by the single condition
\begin{equation}
	[G^{\mu\nu}]n_\mu n_\nu = 0\label{RH2}
\end{equation}
so long as the shock surface is not null and the areas of the spheres of symmetry match smoothly at the shock and change monotonically as the shock evolves. Note that in general, assuming that the angular variables are identified across the shock, we expect conservation to entail two conditions, one for the time and one for the radial components. The fact that the smooth matching of the spheres of symmetry reduces conservation to one condition can be interpreted as an instance of the general principle that directions of smoothness in the metric imply directions of conservation of the sources.
\begin{theorem}
\label{T4}
	Assume that $g$ and $\bar{g}$ are two spherically symmetric metrics that match Lipschitz continuously across a three dimensional shock surface $\Sigma$ to form the matched metric $g\cup\bar{g}$. That is, assume that $g$ and $\bar{g}$ are Lorentzian metrics given respectively by
	\begin{align}
		ds^2 &= -a(t,r)dt^2 + b(t,r)dr^2 + c(t,r)d\Omega^2,\\
		d\bar{s}^2 &= -\bar{a}(\bar{t},\bar{r})d\bar{t}^2 + \bar{b}(\bar{t},\bar{r})d\bar{r}^2 + \bar{c}(\bar{t},\bar{r})d\Omega^2,
	\end{align}
	and that there exists a smooth coordinate transformation $\Psi : (t,r) \rightarrow (\bar{t},\bar{r})$, defined in a neighbourhood of a shock surface $\Sigma$ given by $r=r(t)$, such that the metrics agree on $\Sigma$ (with the implicit assumption that $\theta$ and $\varphi$ are identified). Moreover, assume that
	\begin{equation}
		c(t,r) = \bar{c}(\Psi(t,r)),
	\end{equation}
	in an open neighbourhood of the shock surface $\Sigma$, so that, in particular, the areas of the two-spheres of symmetry in the barred and unbarred metrics agree at the shock surface. Furthermore, assume that the shock surface $r=r(t)$ in unbarred coordinates is mapped to the surface $\bar{r} = \bar{r}(\bar{t})$ by $(\bar{t},\bar{r}(\bar{t})) = \Psi(t,r(t))$, that the normal $\boldsymbol{n}$ to $\Sigma$ is non-null, and that
	\begin{equation}
		\boldsymbol{n}(c)\neq0
	\end{equation}
	where $\boldsymbol{n}(c)$ denotes the derivative of $c$ in the direction of the vector $\boldsymbol{n}$.\footnotemark[7]\footnotetext[7]{That is, we assume that the areas of the two-spheres of symmetry change monotonically in the direction normal to the surface. For example, if $c = r^2$, then $\frac{\partial c}{\partial t} = 0$, so the assumption $\boldsymbol{n}(c) \neq 0$ is valid except when $\boldsymbol{n} = \frac{\partial}{\partial t}$, in which case the rays of the shock surface would be spacelike. Thus the shock speed would be faster than the speed of light if our assumption $\boldsymbol{n}(c) \neq 0$ failed in the case $c=r^2$.} Then the following are equivalent:
	\begin{enumerate}
		\item The components of the metric $g\cup\bar{g}$ in any Gaussian normal coordinate system are $C^{1,1}$ functions of these coordinates across the surface $\Sigma$.
		\item $[G^{\mu\nu}]n_\mu = 0$.
		\item $[G^{\mu\nu}]n_\mu n_\nu = 0$.
		\item $[K]=0$.
	\end{enumerate}
	where $[f]=\bar{f}-f$ denotes the jump in the quantity $f$ across $\Sigma$, and $K$ is the second fundamental form on the shock surface.
\end{theorem}

\section{Shock-Wave Solutions Inside the Hubble Radius - The Case $r_*=0$}
\label{S4}

To construct the family of shock-wave solutions for parameter values $0 < \sigma \leq 1$ and $r_*=0$, we match the explicitly defined solution (\ref{FRW3})-(\ref{FRW5}) of the FRW metric (\ref{FRW}) to the explicitly given TOV metric (\ref{TOV}) outside the black hole, that is, with $A>0$. In this case, we can bypass the problem of deriving and solving the ODE for the shock surface and constraints discussed above, by actually deriving the explicit solution of the Einstein-Euler equations of TOV form that meets these equations. This explicitly defined solution represents the general relativistic version of a static singular isothermal fluid sphere, that is, singular because it has an inverse square density profile and isothermal because the relationship between the density and pressure is $\bar{p} = \bar{\sigma}\bar{\rho}$ with constant $\sigma$.

Assuming the stress-energy-momentum tensor for a perfect fluid and assuming that the density and pressure depend only on $\bar{r}$, the Einstein-Euler equations for the TOV metric (\ref{TOV}) outside the black hole are equivalent to the Oppenheimer--Volkoff system:
\begin{align}
	\frac{dM}{d\bar{r}} &= 4\pi\bar{r}^2\bar{\rho},\label{OV1}\\
	-\bar{r}^2\frac{d\bar{p}}{d\bar{r}} &= \mathcal{G}M\bar{\rho}\left(1+\frac{\bar{p}}{\bar{\rho}}\right)\left(1+\frac{4\pi\bar{r}^3\bar{p}}{M}\right)\left(1-\frac{2\mathcal{G}M}{\bar{r}}\right)^{-1}.\label{OV2}
\end{align}
Integrating (\ref{OV1}) we obtain the usual interpretation of $M$ as the total mass inside radius $\bar{r}$,
\begin{equation}
	M(\bar{r}) = \int_{0}^{\bar{r}}4\pi\xi^2\bar{\rho}(\xi)d\xi.\label{M3}
\end{equation}
The metric component $B$ is determined from $\bar{\rho}$ and $M$ through the equation
\begin{equation}
	\frac{1}{B}\frac{dB}{d\bar{r}} = -\frac{2}{\bar{p}+\bar{\rho}}\frac{d\bar{p}}{d\bar{r}}.\label{OV3}
\end{equation}
Assuming
\begin{align}
	\bar{p} &= \bar{\sigma}\bar{\rho},\label{OV4}\\
	\bar{\rho} &= \frac{\gamma}{\bar{r}^2},\label{OV5}
\end{align}
for some constants $\bar{\sigma}$ and $\gamma$, then substituting this into (\ref{M3}), we obtain
\begin{equation}
	M(\bar{r}) = 4\pi\gamma\bar{r}.\label{M4}
\end{equation}
Putting (\ref{OV4})-(\ref{M4}) into (\ref{OV2}) and simplifying
yields the identity
\begin{equation}
	\gamma = \frac{1}{2\pi\mathcal{G}}\left(\frac{\bar{\sigma}}{1+6\bar{\sigma}+\bar{\sigma}^2}\right).\label{gamma}
\end{equation}
From (\ref{M3}) we obtain
\begin{equation}
	A = 1 - 8\pi\mathcal{G}\gamma < 1.\label{M5}
\end{equation}
Applying (\ref{OV3}) leads to
\begin{equation}
	B = B_0\left(\frac{\bar{\rho}}{\bar{\rho}_0}\right)^{-\frac{2\bar{\sigma}}{1+\bar{\sigma}}} = B_0\left(\frac{\bar{r}}{\bar{r}_0}\right)^{\frac{4\bar{\sigma}}{1+\bar{\sigma}}}.\label{OV6}
\end{equation}
By rescaling the time coordinate, we can take $B_0 = 1$ at $\bar{r}_0 = 1$, in which case (\ref{OV6}) reduces to
\begin{equation}
	B = \bar{r}^{\frac{4\bar{\sigma}}{1+\bar{\sigma}}}.\label{OV7}
\end{equation}
We conclude that when (\ref{gamma}) holds, (\ref{OV4})-(\ref{M5}) and (\ref{OV6}) provide an explicit solution of the Einstein-Euler equations of TOV type for each $0 < \bar{\sigma} \leq 1$.\footnotemark[8]\footnotetext[8]{In this case, an explicit solution of TOV type was first found by Tolman \cite{T1939} and rediscovered in the case $\bar{\sigma} = \frac{1}{3}$ by Misner and Zapolsky (\cite{W1972}, page 320).} By (\ref{M5}), these solutions are defined outside the black hole since $\bar{r} > 2M$. When $\bar{\sigma} = \frac{1}{3}$, (\ref{gamma}) yields $\gamma = \frac{3}{56\pi{\cal G}}$ (\cite{W1972}, equation (11.4.13)).

To match the explicitly given FRW solution (\ref{FRW3})-(\ref{FRW5}) with equation of state $p = \sigma\rho$ to the explicitly given TOV solution (\ref{OV4})-(\ref{OV7}) with equation of state $\bar{p} = \bar{\sigma}\bar{\rho}$ across a shock interface, we first set $\bar{r} = Rr$ to match the spheres of symmetry and then match the timelike and spacelike components of the corresponding metrics in standard Schwarzschild coordinates. The matching of the $d\bar{r}^2$ coefficients yields the conservation of mass condition that implicitly specifies the shock surface $\bar{r} = \bar{r}(t)$,
\begin{equation}
	M(\bar{r}) = \frac{4\pi}{3}\rho(t)\bar{r}^3.
\end{equation}
Using this together with (\ref{OV5}) and (\ref{M4}) gives the following two relations that hold at the shock surface:
\begin{align}
	\bar{r} &= \sqrt{\frac{3\gamma}{\rho(t)}},\\
	\rho &= \frac{3}{4\pi}\frac{M}{\bar{r}(t)^3} = \frac{3\gamma}{\bar{r}(t)^2} = 3\bar{\rho}.
\end{align}
Matching the $d\bar{t}^2$ coefficients on the shock surface determines the integrating factor $\psi$ (see Section \ref{S2}) in a neighbourhood of the shock surface by assigning initial conditions for (\ref{FRWSchwarz2}). Finally, the conservation constraint $[T^{\mu\nu}]n_\mu n_\nu=0$ leads to the single condition
\begin{equation}
	(1-A)(\rho+\bar{p})(p+\bar{\rho})^2 + \left(1-\frac{1}{A}\right)(\bar{\rho}+\bar{p})(\rho+p)^2 + (p-\bar{p})(\rho-\bar{\rho})^2 = 0,
\end{equation}
which upon using $p = \sigma\rho$ and $\bar{p} = \bar{\sigma}\bar{\rho}$ is satisfied providing $\sigma$ and $\bar{\sigma}$ are related by
\begin{equation}
	\bar{\sigma} = \frac{1}{2}\sqrt{9\sigma^2+54\sigma+49} - \frac{3}{2}\sigma - \frac{7}{2} =: H(\sigma).\label{barsigma}
\end{equation}
Alternatively, we can solve for $\sigma$ in (\ref{barsigma}) and write this relation as
\begin{equation}
	\sigma = \frac{\bar{\sigma}(\bar{\sigma}+7)}{3(1-\bar{\sigma})}.
\end{equation}
This guarantees that conservation holds across the shock surface. It thus follows from Theorem \ref{T4} that all of the equivalencies in Theorem \ref{T3} hold across the shock surface. Note that $H(0)=0$ and to leading order
\begin{equation}
	\bar{\sigma}=\frac{3}{7}\sigma+O(\sigma^2)
\end{equation}
as $\sigma\rightarrow0$. Within the physical region $0 \leq \sigma,\bar{\sigma} \leq 1$, $H'(\sigma)>0$, $\bar{\sigma} < \sigma$ and
\begin{align*}
	H\left(\frac{1}{3}\right) &= \sqrt{17}-4 \approx 0.1231,\\
	H(1) &= \sqrt{28}-5 \approx 0.2915.
\end{align*}
Using the formulas for the FRW metric in (\ref{FRW3})-(\ref{FRW5}) and setting $R_0=1$ at $\rho = \rho_0$, $t=t_0$, we obtain the following formulas for the shock position:
\begin{align}
	\bar{r}(t) &= \alpha t,\label{SP1}\\
	r(t) &= \frac{\bar{r}(t)}{R(t)} = \beta t^{\frac{1+3\sigma}{3+3\sigma}},\label{SP2}
\end{align}
where
\begin{align}
	\alpha &= 3(1+\sigma)\sqrt{\frac{\bar{\sigma}}{1+6\bar{\sigma}+\bar{\sigma}^2}},\label{alpha}\\
	\beta &= \alpha^{\frac{1+3\sigma}{3+3\sigma}}\left(\frac{3\gamma}{\rho_0}\right)^{\frac{1}{3+3\sigma}}.\label{beta}
\end{align}
It follows from (\ref{M4}) that $A>0$ and from (\ref{SP2}) that $r_*=\lim_{t\rightarrow0}r(t)=0$. The entropy condition that the shock wave be compressive follows from the fact that $\bar{\sigma} = H(\sigma) < \sigma$. Thus we conclude that for each $0 < \sigma \leq 1$, $r_*=0$, the solutions constructed in (\ref{OV4})-(\ref{beta}) define a one-parameter family of shock-wave solutions that evolve everywhere outside the black hole, which implies that the distance from the shock wave to the FRW centre is less than one Hubble length for all $t>0$. Using (\ref{SP1}) and (\ref{SP2}), one can determine the shock speed and check when the Lax characteristic condition \cite{L1957} holds at the shock. The result is the following (see \cite{ST1995,A2022} for details).\footnotemark[9]\footnotetext[9]{Note that even when the shock speed is larger than $c$, only the wave, and not the sound speeds or any other physical motion, exceeds the speed of light.}
\begin{theorem}
	Let
	\begin{align*}
		\sigma_1 &= \frac{\sqrt{10}+1}{9} \approx 0.462,\\
		\sigma_2 &= \frac{\sqrt{5}}{3} \approx 0.745,
	\end{align*}
	then the Lax characteristic condition holds at the shock if and only if $0 < \sigma \leq \sigma_1$ and the shock speed is subluminal (less than the speed of light) if and only if $ 0 < \sigma < \sigma_2$.
\end{theorem}
The explicitly defined solution in the case $r_*=0$ can be interpreted as the general relativistic version of a shock-wave explosion into a static singular isothermal fluid sphere, known in the Newtonian case as a simple model for star formation \cite{ST1998B}. As the scenario goes, a star begins as a diffuse cloud of gas. The cloud slowly contracts under its own gravitational force by radiating energy out as gravitational potential energy is converted into kinetic energy. This contraction continues until the gas cloud reaches the point where the mean free path for transmission of light is small enough that light is scattered, instead of transmitted, through the cloud. The scattering of light within the gas cloud has the effect of equalising the temperature within the cloud, and at this point the gas begins to drift toward the most compact configuration of the density that balances the pressure when the equation of state is isothermal. This configuration is a static singular isothermal sphere, the general relativistic version of which is the explicitly given TOV solution beyond the shock wave when $r_*=0$. This solution in the Newtonian case is also inverse square in the density and pressure and so the density tends to infinity at the centre of the sphere. Eventually, the high density at the centre ignites a thermonuclear reaction. The result is a shock-wave explosion emanating from the centre of the sphere, with this explosion signifying the birth of a star. The solutions when $r_*=0$ represent the general relativistic version of such a shock-wave explosion.

\section{Shock-Wave Solutions Beyond the Hubble Radius - The Case $r_*>0$}
\label{S5}

When the shock wave is beyond one Hubble length from the FRW centre, we obtain a family of shock-wave solutions for each $0 < \sigma \leq 1$ and $r_*>0$ by matching the FRW metric (\ref{FRW}) to a TOV metric of the form (\ref{TOV}) to form a shock wave under the assumption that
\begin{align}
	A(\bar{r}) = 1 - \frac{2M(\bar{r})}{\bar{r}} =: 1 - N(\bar{r}) < 0.
\end{align}
In this case, $\bar{r}$ is the timelike variable. Assuming the stress-energy-momentum tensor $T$ is taken to be that of a perfect fluid co-moving with the TOV metric, the Einstein equations $G = \kappa T$ (inside the black hole) take the form \cite{ST2004},
\begin{align}
	\bar{p}' &= \frac{\bar{p}+\bar{\rho}}{2}\frac{N'}{N-1},\label{IBH1}\\
	N' &= -\left(\frac{N}{\bar{r}} + \kappa\bar{p}\bar{r}\right),\label{IBH2}\\
	\frac{B'}{B} &= -\frac{1}{N-1}\left(\frac{N}{\bar{r}} + \kappa\bar{\rho}\right).\label{IBH3}
\end{align}
The system (\ref{IBH1})-(\ref{IBH3}) defines the simplest class of gravitational metrics that contain matter, evolve inside the black hole and such that the mass function $M(\bar{r}) < \infty$ at each fixed time $\bar{r}$. System (\ref{IBH1})-(\ref{IBH3}) for $A<0$ differs substantially from the TOV equations for $A>0$ because, for example, the energy density $T^{00}$ is equated with the timelike component $G^{rr}$ when $A<0$ but with $G^{tt}$ when $A>0$. In particular, this implies that inside the black hole the mass function $M(\bar{r})$ does not have the interpretation as the total mass inside radius $\bar{r}$ as it does outside the black hole.

Equations (\ref{IBH2}), (\ref{IBH3}) do not have the same character as (\ref{OV1}), (\ref{OV2}), and the relation $\bar{p} = \bar{\sigma}\bar{\rho}$ with constant $\sigma$ is inconsistent with (\ref{IBH2}), (\ref{IBH3}) together with the conservation constraint and the FRW assumption $p = \sigma\rho$ for shock-wave matching. Thus, instead of looking for an explicit solution of (\ref{IBH2}), (\ref{IBH3}) ahead of time, as in the case $r_*=0$, we assume the FRW solution (\ref{FRW3})-(\ref{FRW5}) and derive the ODE that describe the TOV metrics that match this FRW metric Lipschitz continuously across a shock surface and then impose the conservation, entropy and equation of state constraints at the end. Matching a given flat ($k=0$) FRW metric to a TOV metric inside the black hole across a shock interface leads to the system of ODE \cite{ST2004},
\begin{align}
	\frac{du}{dN} &= -\left(\frac{(1+u)}{2(1+3u)N}\right)\left(\frac{(3u-1)(\sigma-u)N+6u(1+u)}{(\sigma-u)N+(1+u)}\right),\label{ueqn}\\
	\frac{d\bar{r}}{dN} &= -\frac{1}{1+3u}\frac{\bar{r}}{N},\label{Neqn}
\end{align}
with conservation constraint
\begin{align}
	w = \frac{(\sigma-u)N-\sigma(1+u)}{(\sigma-u)N+(1+u)},\label{weqn}
\end{align}
where
\begin{align}
	u &= \frac{\bar{p}}{\rho}, & w &= \frac{\bar{\rho}}{\rho}, & \sigma &= \frac{p}{\rho}.
\end{align}
Here $\rho$ and $p$ denote the (known) FRW density and pressure and all variables are evaluated at the shock. Solutions of (\ref{ueqn})-(\ref{weqn}) determine the (unknown) TOV metrics that match the given FRW metric Lipschitz continuously across a shock interface such that conservation of energy and momentum hold across the shock and such that there are no delta function sources at the shock \cite{I1966,ST1997B}. Note that the dependence of (\ref{ueqn})-(\ref{weqn}) on the FRW metric is only through the variable $\sigma$, and so the advantage of taking constant $\sigma$ is that the whole solution is determined by the inhomogeneous scalar equation (\ref{ueqn}) when $\sigma$ is constant. We take as the entropy constraint the condition that
\begin{align}
	0 &< \bar{p} < p, & 0 &< \bar{\rho} < \rho,\label{Entropy}
\end{align}
and to insure a physically reasonable solution, we impose the equation of state constraint on the TOV side of the shock,\footnotemark[10]\footnotetext[10]{This is equivalent to the dominant energy condition \cite{BG1987}.}
\begin{equation}
	0 < \bar{p} < \bar{\rho}.\label{EoS1}
\end{equation}
Condition (\ref{Entropy}) implies that outgoing shock waves are compressive. Inequalities (\ref{Entropy}) and (\ref{EoS1}) are both implied by the single condition \cite{ST2004},
\begin{equation}
	\frac{1}{N} < \left(\frac{1-u}{1+u}\right)\left(\frac{\sigma-u}{\sigma+u}\right).\label{EoS2}
\end{equation}
Since $\sigma$ is constant, equation (\ref{ueqn}) uncouples from (\ref{Neqn}), and thus solutions of system (\ref{ueqn})-(\ref{weqn}) are determined by the scalar non-autonomous equation (\ref{ueqn}). Making the change of variable $S = \frac{1}{N}$, which transforms the Big Bang $N\rightarrow\infty$ over to a rest point at $S\rightarrow0$, we obtain,
\begin{equation}
	\frac{du}{dS} = \left(\frac{(1+u)}{2(1+3u)S}\right)\left(\frac{(3u-1)(\sigma-u)+6u(1+u)S}{(\sigma-u)+(1+u)S}\right).\label{dudS}
\end{equation}
Note that the conditions $N>1$ and $0 < \bar{p} < p$ restrict the domain of (\ref{dudS}) to the region $0 < u < \sigma < 1$, $0 < S < 1$. The next theorem gives the existence of solutions for $0 < \sigma \leq 1$, $r_*>0$ inside the black hole \cite{ST2003}.
\begin{theorem}
	\label{T6}
	For every $0 < \sigma < 1$ there exists a unique solution $u_{\sigma}(S)$ of (\ref{dudS}) such that (\ref{EoS2}) holds for all $0<S<1$. Moreover,
	\begin{align}
		0 &< u_{\sigma}(S) < \bar{u}, & \lim_{S\rightarrow0}u_{\sigma}(S) &= \bar{u}, & \lim_{S\rightarrow1}\bar{p} &= 0, & \lim_{S\rightarrow1}\bar{\rho} &= 0,\label{Limits}
	\end{align} 
	where
	\begin{equation}
		\bar{u} = \min\left\{\frac{1}{3},\sigma\right\}.\label{EoS3}
	\end{equation}
	Furthermore, for each of the solutions $u_{\sigma}(S)$, the shock position is determined by the solution of (\ref{Neqn}), which in turn is determined uniquely by an initial condition which can be taken to be the FRW radial position of the shock wave at the instant of the Big Bang,
	\begin{equation}
		r_*=\lim_{S\rightarrow0}r(S)>0.
	\end{equation}
\end{theorem}
Concerning the the shock speed, we have the following theorem.
\begin{theorem}
	Let $0 < \sigma < 1$. Then the shock speed $s_{\sigma}(S) := s(u_{\sigma}(S)) < 1$ for all $ 0 < S \leq 1$ if and only if $\sigma < \frac{1}{3}$, that is, the shock speed is subluminal if and only if $\sigma < \frac{1}{3}$.
\end{theorem}
For the shock speed near the Big Bang ($S=0$), we have the following theorem:
\begin{theorem}
	\label{T8}
	The shock speed at the Big Bang ($S=0$) is given by:
	\begin{align}
		&\lim_{S\rightarrow0}s_{\sigma}(S) = 0, & \sigma &< \frac{1}{3},\\
		&\lim_{S\rightarrow0}s_{\sigma}(S) = 1, & \sigma &= \frac{1}{3},\\
		&\lim_{S\rightarrow0}s_{\sigma}(S) = \infty, & \sigma &> \frac{1}{3}.
	\end{align}
\end{theorem}
Theorem \ref{T8} shows that the equation of state $p = \frac{1}{3}\rho$ plays a special role in the analysis when $r_*>0$, and only for this equation of state does the shock wave emerge at the Big Bang at a finite non-zero speed, the speed of light. Moreover, (\ref{EoS3}) implies that in this case, the correct relation $\bar{p} = \bar{\sigma}\bar{\rho}$ is also achieved in the limit $S\rightarrow0$. The result (\ref{Limits}) implies that (neglecting the pressure $p$ at this time onward) the solution continues to a $k=0$ Oppenheimer--Snyder solution outside the black hole for $S>1$.

It follows that the shock wave will first become visible at the FRW centre $\bar{r}=0$ at the moment $t=t_0$ (where $R(t_0)=1$) when the Hubble length $1/H_0 := 1/H(t_0)$ satisfies
\begin{equation}
	\frac{1}{H_0} = \frac{1+3\sigma}{2}r_*,
\end{equation}
where $r_*$ is the FRW position of the shock at the instant of the Big Bang. At this time, the number of Hubble lengths $\sqrt{N_0}$ from the FRW centre to the shock wave at time $t=t_0$ can be estimated by
\begin{equation}
	1 \leq \frac{2}{1+3\sigma} \leq \sqrt{N}_0 \leq \frac{2}{1+3\sigma}e^{\sqrt{3\sigma}\left(\frac{1+3\sigma}{1+\sigma}\right)}.
\end{equation}
Thus, in particular, the shock wave will still lie beyond the Hubble radius $1/H_0$ at the FRW time $t_0$ when it first becomes visible. Furthermore, the time $t_{crit} > t_0$ at which the shock wave will emerge from the white hole, given that $t_0$ is the first instant at which the shock becomes visible at the FRW centre, can be estimated by
\begin{align}
	\frac{2}{1+3\sigma}e^{\frac{1}{4}\sigma} &\leq \frac{t_{crit}}{t_0} \leq \frac{2}{1+3\sigma}e^{\frac{\sqrt{12\sigma}}{1+\sigma}}, & 0 &< \sigma < \frac{1}{3},\label{Estimate1}\\
	e^{\frac{\sqrt{6}}{4}} &\leq \frac{t_{crit}}{t_0} \leq e^{\frac{3}{2}}, & \sigma &= \frac{1}{3}.\label{Estimate2}
\end{align}
Inequalities (\ref{Estimate1}), (\ref{Estimate2}) imply, for example, that at the Oppenheimer--Snyder limit $\sigma=0$:
\begin{align*}
	\sqrt{N_0} &= 2, & \frac{t_{crit}}{t_0} &= 2
\end{align*}
and in the limit $\sigma = \frac{1}{3}$:
\begin{align*}
	1 &< \sqrt{N_0} \leq 4.5, & 1.8 &\leq \frac{t_{crit}}{t_0} \leq 4.5.
\end{align*}
We can conclude that the moment $t=t_0$, when the shock wave first becomes visible at the FRW centre, the shock wave must lie within 4.5 Hubble lengths of the FRW centre. Throughout the expansion up until this time, the expanding universe must lie entirely within a white hole, that is, the Universe will eventually emerge from this white hole, but not until some later time $t_{crit}$, where $t_{crit}$ does not exceed $4.5t_0$.

\section{Self-Similar Extensions of FRW-TOV Shock-Waves Inside the Hubble Radius}
\label{S6}

The previous two sections focused on constructing general relativistic shock waves with the explicitly known flat ($k=0$) FRW metric forming the interior expanding wave. In this section we consider a broader family of expanding waves and demonstrate how to impose conservation across a shock surface with a TOV spacetime on the exterior. This achieves three things: The first is the determination of all expanding waves that are regular at the radial centre ($\bar{r} = 0$) that can be matched with conservation to a TOV spacetime inside the Hubble radius (outside the black hole), the second is the determination of the unique expanding wave with a pure radiation equation of state ($p = \frac{1}{3}\rho$) that can be matched to the pure radiation TOV spacetime, and the third is the determination of the accelerated expansion that results from a general relativistic explosion with a TOV exterior.

Extending the flat FRW spacetime to a family of expanding waves is motivated by the fact that both the FRW and TOV spacetimes are self-similar in the variable $\xi = r/t$, that is, the metric components and hydrodynamic variables depend only on the single variable $\xi$. The self-similarity of the TOV spacetime means that to match an expanding wave to this spacetime with conservation across the shock and with an equation of state of the form $p = \sigma\rho$, it must be the case that the expanding wave also be self-similar in $\xi$ \cite{CT1971}. Furthermore, requiring that the expanding waves have a \emph{regular centre}, leaves only one family of self-similar expanding waves, with these being the one-parameter family of \emph{asymptotically FRW} spacetimes described independently in \cite{CC2000B} and \cite{ST2012}. Thus it is the case that the family of spherically symmetric self-similar perturbations of the FRW metric account for all physically admissible expanding waves that can be matched to a self-similar TOV metric with conservation across the shock.

Any self-similar metric may be written, without loss of generality, in the self-similar Schwarzschild coordinate form
\begin{align}
	ds^2 = -B(\xi)dt^2 + \frac{1}{A(\xi)}dr^2 + r^2[d\theta^2 + \sin^2(\theta) d\phi^2].
\end{align}
Under the assumption of spherical symmetry, the fluid four-velocity may also be written, without loss of generality, as $\boldsymbol{u} = (u^0,u^1,0,0)$. Under the normalisation condition $g(\boldsymbol{u},\boldsymbol{u}) = -1$, the fluid four-velocity has only one independent component and can thus be fully specified through the \emph{Schwarzschild coordinate velocity}, defined by
\begin{align}
	v = \frac{1}{\sqrt{AB}}\frac{u^1}{u^0}.
\end{align}
Together with $A$, $B$, $\rho$ and $p$, the Schwarzschild coordinate velocity $v$ is one of five unknown variables that completely specify a solution to the self-similar Einstein-Euler equations. As there are only four independent components of the spherically symmetric Einstein-Euler equations, a barotropic equation of state of the form $p = p(\rho)$ is used to close the system. However, spherical symmetry and self-similarity in the variable $\xi$ restrict this equation of state to the form $p = \sigma\rho$ for constant $\sigma$ \cite{CT1971}. Following the development of Smoller and Temple in \cite{ST2012}, substituting the metric ansatz and equation of state into the Einstein-Euler equations yields the system of nonlinear ODE:
\begin{align}
	\xi\frac{dA}{d\xi} &= -\frac{(3+3\sigma)(1-A)v}{\{\cdot\}_S},\label{dA}\\
	\xi\frac{dG}{d\xi} &= -G\left[\left(\frac{1-A}{A}\right)\frac{(3+3\sigma)[(1+v^2)G-2v]}{2\{\cdot\}_S} - 1\right],\label{dG}\\
	\xi\frac{dv}{d\xi} &= -\left(\frac{1-v^2}{2\{\cdot\}_D}\right)\left[3\sigma\{\cdot\}_S + \left(\frac{1-A}{A}\right)\frac{(3+3\sigma)^2\{\cdot\}_N}{4\{\cdot\}_S}\right],\label{dv}
\end{align}
in addition to the constraint
\begin{align}
	\rho = \frac{3(1-v^2)(1-A)G}{\kappa r^2\{\cdot\}_S}.\label{Constaint}
\end{align}
The variable $G$, not to be confused with the Einstein curvature tensor, is defined by
\begin{align}
	G = \frac{\xi}{\sqrt{AB}}
\end{align}
and the bracketed terms are given (using the form found in \cite{A2022}) by:
\begin{align*}
	\{\cdot\}_S &= 3(G-v) - 3\sigma v(1-Gv),\\
	\{\cdot\}_N &= -3(G-v)^2 + 3\sigma v^2(1-Gv)^2,\\
	\{\cdot\}_D &= \frac{3}{4}(3+3\sigma)\left[(G-v)^2 - \sigma(1-Gv)^2\right].
\end{align*}
Thus a solution is specified fully by the three variables $A$, $G$ and $v$.

The asymptotically FRW spacetimes are described by the family of solutions to (\ref{dA})-(\ref{Constaint}) with the leading order form as $\xi\to 0$ \cite{ST2012}:
\begin{align*}
	A(\xi) &\approx 1 - \frac{1}{4}a^2\xi^2 + O(\xi^4),\\
	G(\xi) &\approx \frac{1}{4}(3+3\sigma)\xi + O(\xi^3),\\
	v(\xi) &\approx \frac{1}{2}\xi + O(\xi^3).
\end{align*}
The parameter $a$ is referred to as the \emph{acceleration parameter}, as changing this parameter value away from $a=1$, which corresponds to the unperturbed FRW spacetime, changes the accelerated expansion of the spacetime as measured by an observer at the radial centre. This change is specified through the modified red-shift versus luminosity relation
\begin{align}
	d_l = 2ct_0\left(z + \frac{a^2-1}{2}z^2 + \frac{(a^2-1)(5a^2+4)}{10}z^3 + |a-1|O(z^4)\right)\label{dl}
\end{align}
where $d_l$ is the luminosity distance, $t_0$ is the time of observation of the radiation and $z$ is the redshift factor \cite{ST2012,A2022}.\footnotemark[11]\footnotetext[11]{Note that (\ref{dl}) was originally derived in \cite{ST2012}, but due to an error in the original paper this expression was corrected and reproduced in \cite{A2022}.} It is also important to note that $a\neq 1$ also breaks the spacial homogeneity of the spacetime, with this change becoming more apparent for larger $|a-1|$ and less apparent closer to $\xi = 0$.

The asymptotically FRW spacetimes are not known explicitly away from $\xi = 0$, instead these solutions are described by the system of ODE (\ref{dA})-(\ref{dv}) and constraint (\ref{Constaint}). In order to match an asymptotically FRW spacetime to a TOV spacetime with conservation holding across the shock, the following lemma is required \cite{A2022}.
\begin{lemma}
	\label{L1}
	Let $(A,G,v)$ denote a solution to (\ref{dA})-(\ref{dv}) and suppose there exists a $\xi_0 > 0$ such that
	\begin{align}
		A(\xi_0) = 1 - 2M(\bar{\sigma}).\label{M6}
	\end{align}
	Then $(A,G,v)$ can be matched to the TOV spacetime (with equation of state $\bar{p} = \bar{\sigma}\bar{\rho}$) on the surface $\xi = \xi_0$ and the Rankine--Hugoniot jump condition is given by
	\begin{align}
		\frac{[\sigma+v^2(\xi_0)]G(\xi_0)-(1+\sigma)G^2(\xi_0)v(\xi_0)}{[1+\sigma v^2(\xi_0)]G(\xi_0)-(1+\sigma)v(\xi_0)} = \bar{\sigma}.\label{RH3}
	\end{align}
\end{lemma}
Thus if a solution satisfies (\ref{M6}) and (\ref{RH3}), then by Lemma \ref{L1} this solution is a shock-wave solution. However, for this shock wave to be stable in the gas-dynamical sense, that is, for the fluid characteristics to impinge on the shock surface from both sides, the following theorem is needed \cite{A2022}. This theorem establishes the \emph{Lax entropy conditions}, also known as the \emph{Lax characteristic conditions}.
\begin{theorem}
	\label{T9}
	Let $(A,G,v)$ denote a solution to (\ref{dA})-(\ref{dv}). If there exists a $\xi_0 > 0$ such that $(A,G,v)$ can be matched to the TOV spacetime (with equation of state $\bar{p} = \bar{\sigma}\bar{\rho}$) to form a shock-wave solution with a subluminal shock speed $(G(\xi_0) < 1)$, then the Lax characteristic conditions are satisfied if:
	\begin{enumerate}
		\item $\sigma = \bar{\sigma}$, or
		\item $\sigma < \bar{\sigma}$ and $G(\xi_0) > \sqrt{\bar{\sigma}}$, or
		\item $\sigma > \bar{\sigma}$ and $\{\cdot\}_D(\xi_0) < 0$.
	\end{enumerate}
\end{theorem}
The addition of the acceleration parameter $a$ means it is possible to specify both the FRW and TOV equation of state independently, with the Rankine--Hugoniot jump condition then determining $a$. We see from Theorem \ref{T9} that if we were to require both equations of state to model pure radiation ($p = \frac{1}{3}\rho$), that is, a uniform equation of state across the shock surface, then the entropy condition is automatically implied from the Rankine--Hugoniot jump condition. Indeed, numerical approximations of such a solution imply $a \approx 2.58$ \cite{A2022}. Such a solution, if taken as a cosmological model, would yield an accelerated expansion many orders of magnitude larger than what is currently observed, since the expected value of $a$ in such a model would be $a \approx 1$ \cite{ST2012}. However, because this model uses the TOV spacetime outside the black hole, the shock surface would be within the Hubble radius and thus expected to be visible at the present time. It remains an active area of research for Alexander and Temple to construct shock-wave solutions with shock surfaces beyond the Hubble radius and determine the accelerated expansion exhibited by these spacetimes. For the solutions constructed beyond the Hubble radius in Section \ref{S5}, it is intriguing that both equations of state tend to $p = \frac{1}{3}\rho$ in the limit of the Big Bang, further reinforcing the expectation that $a\approx 1$. We finish with the following theorem from \cite{A2022}.
\begin{theorem}
	There exists an $a > 1$ such that an asymptotically FRW spacetime can be matched to a TOV spacetime within the Hubble radius to form a pure radiation general relativistic shock wave that satisfies the Lax characteristic conditions.
\end{theorem}

\section{Conclusion}
\label{S7}

We have delved into the mathematics behind general relativistic shock waves that admit asymptotically FRW spacetimes as the expanding wave behind the shock. By placing a TOV spacetime on the exterior, these shock waves model the general relativistic analogue of an explosion within a static singular isothermal fluid sphere. Whether such modifications to the FRW spacetime could provide an alternative to the Standard Model of Cosmology first depends on whether the shock surface lies within the Hubble radius. If it does, such a shock wave should be presently observable, which is not the case. This means that the TOV spacetime on the exterior must be modified to be within a black hole. We have seen that it is possible to construct a shock-wave beyond the Hubble radius through this modification. Such a model has the notable property of containing finite total mass but does not account for the accelerated expansion observed in our Universe today. To introduce an accelerated expansion, it is necessary to extend the flat FRW spacetime to a family of self-similar perturbations. This extension provides an additional free parameter $a$, which allows for the equation of state to be specified independently each side of the shock, and in particular, allows both equations of state to model pure radiation. Such a shock wave would thus be applicable in the Radiation Dominated Epoch of the Early Universe.

Constructing general relativistic shock waves within and beyond the Hubble radius with the explicitly known flat FRW spacetime behind the shock was considered in Sections \ref{S4} and \ref{S5} respectively. The extension of Section \ref{S4} to using self-similar perturbations of the flat FRW spacetime was considered in Section \ref{S6} and permitted the construction of general relativistic shock waves that induce an accelerated expansion. Section \ref{S6} resolved the picture within the Hubble radius, an area which has been in active research since Cahill and Taub's seminal paper on self-similar solutions in General Relativity \cite{CT1971}. The extension of Section \ref{S5} to considering self-similar perturbations of the flat FRW spacetime behind a shock lying beyond the Hubble radius remains an active area of research for Alexander and Temple.\footnotemark[12]\footnotetext[12]{In addition, recent studies of self-similar expanding waves have led to a new characterisation of the instability of the flat FRW spacetime in \cite{STV2017}. Alexander and Temple are currently working on a comprehensive extension of this theory.} If the accelerated expansion induced by this shock wave matches what observations suggest the rate was during the Radiation Dominated Epoch, then such a model would offer a mathematically independent mechanism for the accelerated expansion observed today without the need for a cosmological constant, and thus, without the need for dark energy.

\end{document}